# Understanding the Effect of Market Risks on New Pension System and Government Responsibility

Sourish Das[1], Bikramaditya Datta[2], Shiv Ratan Tiwari[3]

**Abstract**

This study examines how market risks impact the sustainability and performance of the New Pension System (NPS). NPS relies on defined contributions from both employees and employers to build a corpus during the employee's service period. Upon retirement, employees use the corpus fund to sustain their livelihood. A critical concern for individuals is whether the corpus will grow sufficiently to be sustainable or if it will deplete, leaving them financially vulnerable at an advanced age. We explore the impact of market risks on the performance of the corpus resulting from the NPS. To address this, we quantify market risks using Monte Carlo simulations with historical data to model their impact on NPS. We quantify the risk of pension corpus being insufficient and the cost to the Government to hedge the risk arising from guaranteeing the pension.

---

[1] Chennai Mathematical Institute; email: sourish@cmi.ac.in

[2] Indian Institute of Technology, Kanpur; email: bikramd@iitk.ac.in

[3] Indian Institute of Technology, Kanpur; email: shivt@iitk.ac.in

**Background and Motivation**

The National Pension System (NPS), adopted in 2003 [NSDL, 2024], was designed to address fiscal pressures associated with the defined benefits system, which, despite being popular among government employees, posed significant long-term financial risks to the government. The shift to the NPS introduced a defined contributions scheme, transferring risk from the government to the employees. In particular, this transition brought uncertainty regarding pension benefits. The key sources of risks under the defined contributions scheme include risks related to market returns and annuity rates at the time of retirement. These factors can significantly impact the corpus's ability to sustain an individual's livelihood post-retirement, necessitating a thorough investigation into their effects to ensure the system's sustainability and adequacy.

In this paper, we build a quantitative framework that helps us understand how the risks related to various factors such as market returns and annuity rates at retirement affect the retirement corpus and the standard of living post-retirement. Specifically, we consider a representative employee who works for a fixed number of years. Each year the employee contributes a fixed percentage of her salary towards a pension corpus along with a matching share by the employer (government). The corpus is invested in various market instruments, and the corpus grows each year according to the prevailing (stochastic) market rates. We simulate the market rates using a Geometric Brownian Motion with parametric values based on performance of the NPS benchmark portfolio for central government employees. We examine the risk due to market returns by examining the distribution of the values of the corpus at the end of the employment period.

After retirement the employee uses the accumulated corpus to buy an annuity at the annuity rate prevailing at that time. The annuity pays out a constant amount for each year that the employee is alive. We examine risks due to different annuity rates by looking at the distribution of the number of years the employee receives less than half of the last drawn salary. This distribution is also important for public policy purposes. Concerns from employees about the uncertainty in the defined contribution system has led to various debates and modifications in the pension system in various states. In 2023, the Andhra Pradesh government [Andhra Pradesh

Guaranteed Pension System Act, 2023] introduced a "Guaranteed Pension System" which guaranteed employees a top-up amount to ensure a guaranteed pension at the rate of fifty percent of the last drawn salary. We use our framework to provide a numerical estimate of what it would cost the government to provide such a guarantee to the representative employee. Obtaining such an estimate is especially important for sound fiscal planning, considering the large number of employees in the government and recent debates about whether such a policy should also be implemented for central government employees [Business Standard, 2024].

The transition from the defined benefits to the NPS has been accompanied by critical analysis of choices in the design of the NPS – see for instance Sane and Thomas (2013). However, there have been relatively few attempts to quantify the risks in the NPS, with the notable exception of Shah (2003). Compared to Shah (2003), we use a more detailed model (Geometric Brownian Motion) for asset market returns, whose parameters are estimated using recent performance of NPS benchmark portfolios. Our analysis of the post-retirement phase is also richer, as we can look at policy implications of changes in annuity rates and its implications for risk management for the government. We also implement the Monte Carlo simulations in R, and the codes are made available at (https://github.com/sourish-cmi/Pension/blob/main/code/code_20240823_sim.R) for interested readers to conduct further study.

**Data and Assumptions**

We analyse the different risks in a simplified setting, which we outline below.

**Time Frame**: Each period corresponds to a year in our analysis. A representative employee starts contributing to the pension fund at time 0 and retires at time $n$. The employee continues to receive pension for $m$ time periods after retirement. For our baseline scenario, we choose $n = 30, m = 20$, corresponding to 30 years of service period and 20 years of life post-retirement.

**Salary**: The employee's salary has two components – basic pay with a starting value of ₹100 and dearness allowance (DA). We assume that the basic pay component has an annual increment of 3 percent. The DA component is assumed to increase at the rate of the previous year's inflation rate.

**Contribution to corpus**: Each period the employee contributes 10 percent of her salary to the corpus. The employer (government) contributes to the corpus each year an amount equal to 14 percent of the employee's salary. These numbers reflect the current contribution rates as applicable to central government employees.

**Inflation**: We simulate inflation values for each of the $n+m$ periods from a Gaussian distribution with mean 4 and standard deviation 1. This choice reflects the current policy of RBI's inflation target regime of 4% with 2% as margin of error, i.e., keep the inflation in the range of 2% to 6%. Our choice ensures that inflation stays in this range 95% of time.

**Market returns**: We analysed data from the Annual Financial Report of the NPS Trust [NPS Trust, 2022-23]. The report provides data on past performance of pension funds for central government employees. As a starting point, we use data from performance of the benchmark index which reflects the composite performance of government securities (G-secs), corporate bonds, equities and money market instruments, aggregated in the ratio of 49, 35, 14 and 2, respectively. Based on this data, the average annualised return over ten years is approximately 9%, with a standard deviation (annualised volatility) of approximately 5%. *This data serves as the foundation for our analysis.* Based on these figures, in the baseline scenario we assume that over the 30 years of employee's service, the NPS will consistently perform with the same average return and volatility. This means that in some years, the NPS might achieve a 16% return, while in others, it might only realize a 3% return. However, over the long term, it is expected to yield an average annual return of 9% with 5% volatility. This is our first assumption regarding asset market return in this study. We will also examine later how changes in these parametric values affect the corpus.

Our second assumption is that all these returns follow Geometric Brownian Motion [4](GBM),

$$S_t = S_1 \, exp\{(\mu - \sigma^2/2)t + \sigma W_t\},$$

where $W_t \sim N(0, t)$, $S_t$ is the value of the benchmark portfolio at time $t$, $S_1$ is the initial value, $\mu$ is average annual return, assumed to be 9% for the baseline scenario and $\sigma$ is volatility, which is assumed to be 5% for the baseline scenario.

Then the return is defined as

$$r_t = log(S_t) - log(S_{t-1}).$$

We require a probability model regarding market returns to make statistical inferences, and these are the only two assumptions made regarding asset market returns throughout the study.

The corpus changes each year due to the previous year's corpus growing at a rate equal to the return simulated above, as well as the new contributions to the corpus.

**Annuity:** At the end of the service period, the entire corpus accumulated is used to buy an annuity, which pays out a fixed amount of pension each year depending on the annuity rate prevailing at the time of purchase of the annuity. For instance, if the corpus amount at the end of the service period is ₹4000 and the annuity rate is 7 percent, then the annuity will pay out ₹280 of pension each year till the employee passes away. There is uncertainty about the annuity rate that might prevail at the time of retirement. In the baseline scenario, we assume an annuity rate equal to 7 percent. We will examine later how changes in the annuity rate affect the standard of post-retirement living.

**Inflation Adjusted Requirement:** One question of interest is whether the pension amount would be sufficient to maintain an adequate standard of living. Given current policy debates, a natural benchmark of such a standard of living is if the pension amount is at least equal or more than 50 percent of the last salary. Given increases in the cost of living due to inflation, we look at the number of years where the pension amount received is less than 50 percent of the last salary, adjusted for inflation.

---

[4] For more details regarding simulations of Geometric Brownian Motion, refer to Das and Sen (2023), and Glasserman (2003).

# Monte Carlo Simulation

Consider a representative employee with a starting basic pay of ₹100 and an annual increment of 3%. The service period is assumed to be 30 years. Each year, a total of 24% of the salary is contributed towards the pension fund, with 10% coming from the employee and 14% from the employer. To model the returns on the pension fund, we use Geometric Brownian Motion (GBM) simulations. These simulations assume an average return of 9% and a standard deviation of 5%. By applying GBM, we can obtain the final corpus amount at the end of the 30 years.

| Year | Inflation | Basic Pay | DA | Salary | Contribution | Returns | Corpus |
|------|-----------|-----------|------|--------|--------------|---------|--------|
| 1    | 1.97%     | 100.00    | 0.00 | 100.00 | 24.00        | 0%      | 24.00  |
| 2    | 2.92%     | 103.00    | 1.97 | 104.97 | 25.19        | 13%     | 52.27  |
| 3    | 3.61%     | 106.09    | 3.01 | 109.10 | 26.18        | 14%     | 85.73  |
| 4    | 4.69%     | 109.27    | 3.83 | 113.10 | 27.14        | 10%     | 121.27 |
| ...  |           |           |      | ...    | ...          | ...     | ...    |
| 30   | 5.46%     | 235.66    | 4.27 | 239.93 | 57.58        | 14%     | 3807.78|

Table 1: One-Sample from GBM. Note that annual returns are random simulations from GBM.

Table 1 presents one simulation from the Geometric Brownian Motion (GBM) model, where returns are generated using the GBM and inflation values are simulated from a Gaussian distribution. The basic pay grows at a fixed rate of 3% per year, and contributions are predefined as discussed above. Over 30 years, the corpus grows to ₹3807.78. However, this is just one random sample from the GBM. We conduct 1,000 such simulations to obtain a comprehensive analysis.

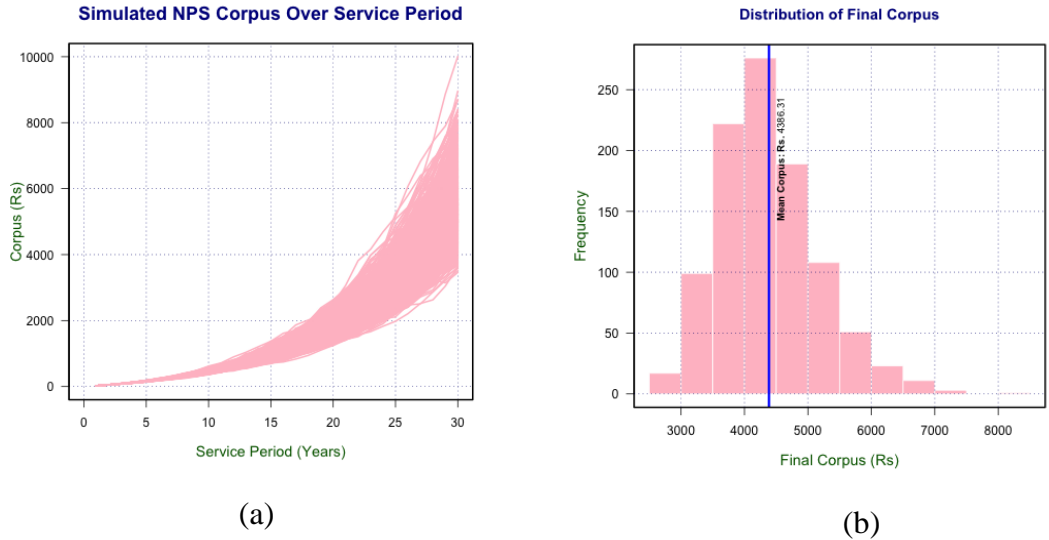

(a)                                                                  (b)

Figure 1: Thousand simulation of the corpus are shown in panel (a). In panel (b) we present the histogram of the value of corpus at the end of service period for $\mu = 9\%$ and $\sigma = 5\%$.

Figure 1 provides an analysis of the corpus of over 1,000 simulations using the Geometric Brownian Motion (GBM) model. In panel (a), the simulations show that the corpus at the end of 30 years could range anywhere between ₹2,730.42 and ₹8,300.07. Panel (b) presents a histogram of the corpus sizes at the end of 30 years from these simulations, revealing that the average corpus size is ₹4386.31, and the standard deviation is ₹792.21. This distribution highlights the variability in the corpus outcomes, emphasising the impact of market volatility on pension fund performance.

We can also use the framework developed to look at how changes in the average annualised returns and standard deviation (annualised volatility) affect the distribution of the final corpus.

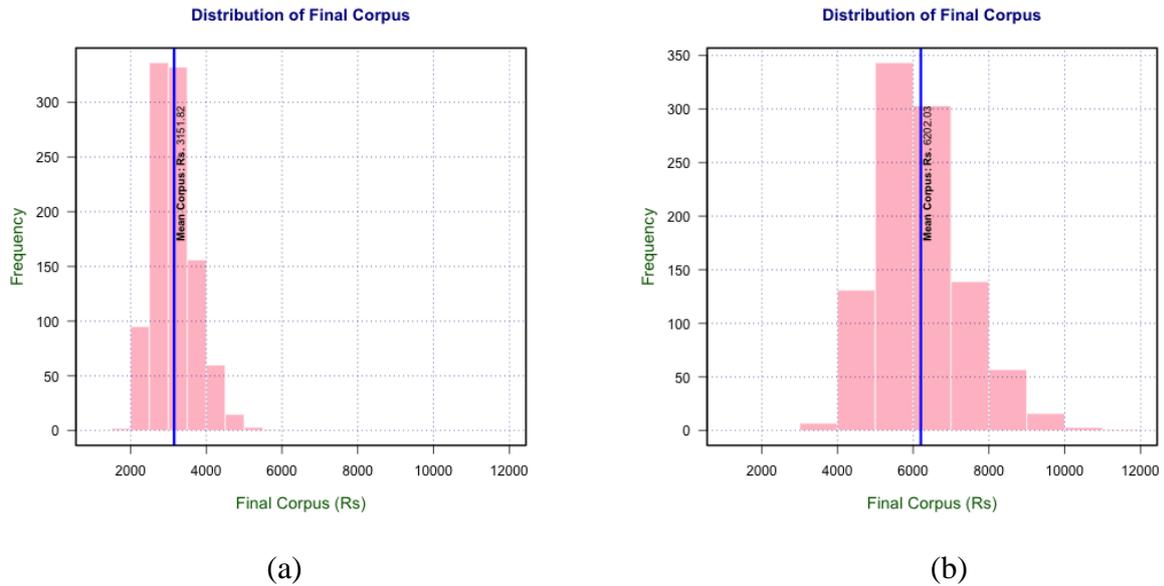

(a)                          (b)

Figure 2: We present the histogram of the value of corpus at the end of service period for $\mu = 7\%$ (panel a) and for $\mu = 11\%$ (panel b). In both cases we considered volatility to be $\sigma = 5\%$.

As expected, a lower value of $\mu$ leads to the distribution of the final corpus shifting to the left with a lower mean value of the final corpus and a lower standard deviation due to compounding effect.

**Retirement Phase**

The corpus at the end of the service period is used to buy an annuity. As part of the baseline scenario, we assume the annuity rate to be equal to 7 percent. This generates a constant annual pension to the employee post-retirement. A point of interest is whether the pension amount would be adequate to sustain a reasonable standard of living. In the context of ongoing policy discussions, a reasonable benchmark for this standard would be if the pension amount is at least 50 percent of the final salary. Considering the rising cost of living due to inflation, we examine the number of years during which the pension amount falls below 50 percent of the final salary when adjusted for inflation.

| Year | Inflation | Inflation Adjusted Requirement | Pension | Sufficient |
|---|---|---|---|---|
| 31 | 5.26 | 126.52 | 266.55 | Yes |
| 32 | 3.88 | 133.17 | 266.55 | Yes |
| 33 | 5.12 | 138.34 | 266.55 | Yes |
| … | … | … | … | |
| … | … | … | … | |
| 50 | 3.33 | 272.85 | 266.55 | No |

Table 2: One sample of pension after retirement

Table 2 presents one simulation of the $m = 20$ years after retirement. The final corpus, worth ₹ 3807.78 was used to purchase annuity at an annuity rate equal to 7 percent. This leads to a yearly pension of ₹266.55. We see that this pension is adequate to cover the inflation adjusted required amount in the initial years of the retirement period. However, in later years it is not sufficient.

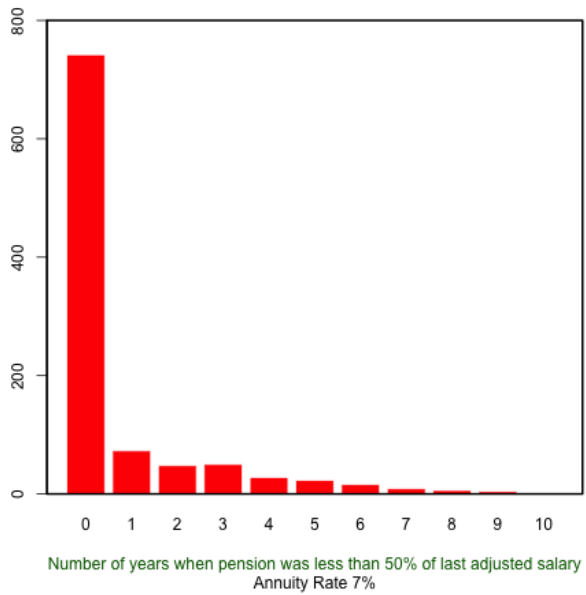 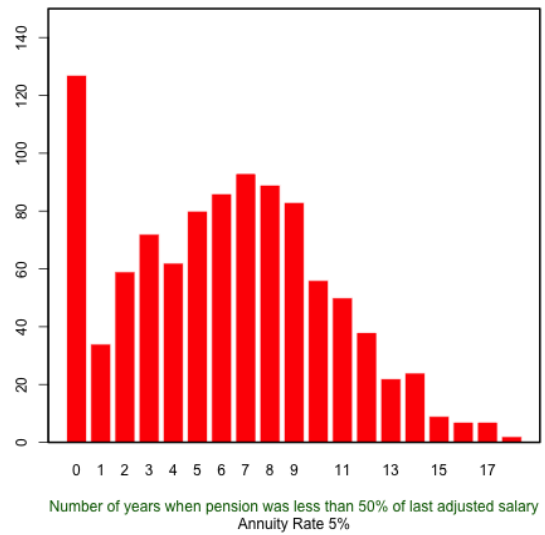

(a)  (b)

Figure 3: Number of years when pension was less than 50% of last adjusted salary with annuity rate = 7% in panel (a) and 5% in panel (b) for 20 years of retirement life.

**Hedging the Risk due to Pension Guarantee**

We can use the framework developed to generate an estimate of the cost of government guarantees. For instance, consider a proposal by the government to guarantee 50 percent of the last salary drawn by the employee at the time of retirement. This means that if the last salary drawn by the employee was ₹250, then the government would intervene in case the pension amount received was less than ₹125 and the government would provide a top-up amount to ensure a guaranteed pension at the rate of fifty percent of the last drawn salary. As before, we adjust the 50 percent by changes in the cost of living, as captured by the changes in inflation rate. Hence, we look at the cost of guaranteeing the inflation adjusted requirement from the perspective of the government.

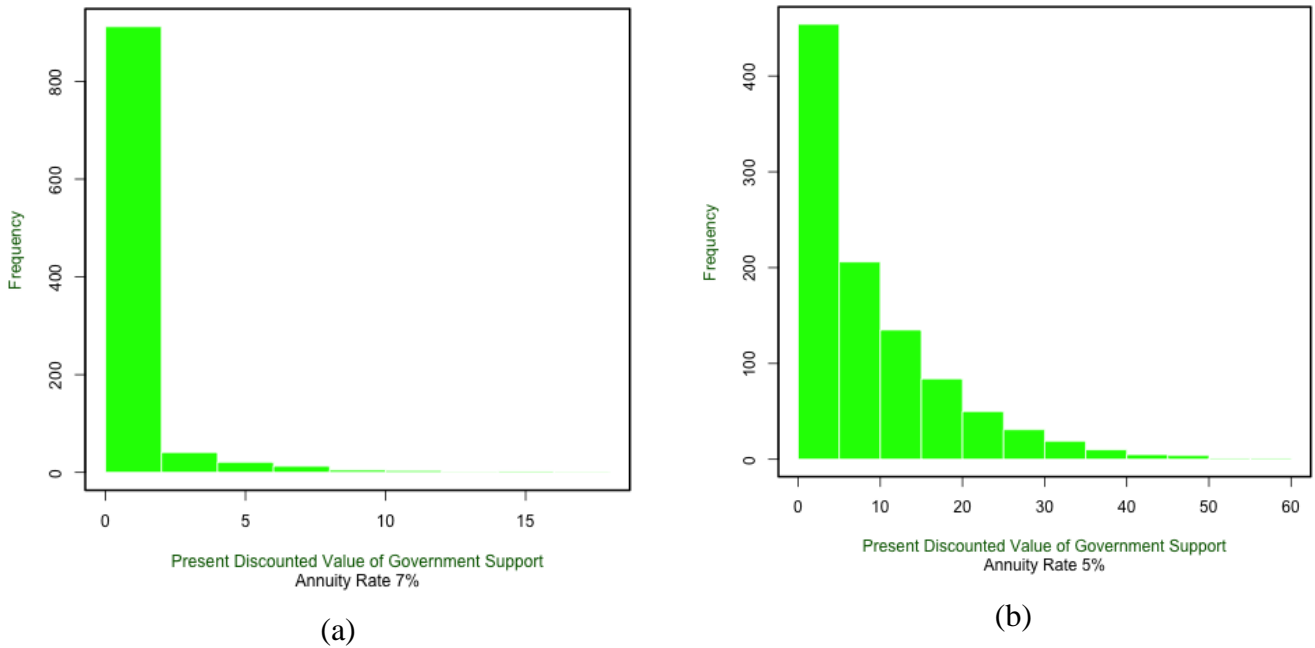

(a)

(b)

Figure 4: Present Discounted Value of Government Support with annuity rate = 7% in panel (a) and 5% in panel (b) for 20 years of retirement life.

Assuming a risk-free rate of 7 percent (a number which approximates the current yield on 30 year Government bond of India) we can calculate the mean present discounted value (in year 1) of providing such guarantees is ₹0.57 when the annuity rate is 7%. However this rises to ₹8.77, when the annuity rate falls to 5%. The above calculation indicates that on average the Government needs to spend 8.77% of employee's first year's basic salary to cover the pension shortage during retirement period. The Government can invest this amount in a long-term bond to mitigate the uncertainty related to the pension guarantee from the market risks.

It is useful to benchmark the above numbers against the ones discussed in the context of new policy developments, for instance, the recently announced Unified Pension Scheme (UPS). The UPS scheme aims to provide retirees pension equal to 50 percent of basic pay earned in the last year of service period (along with adjustments for inflation), and involves the employee contributing 10% of salary (same as the above NPS) each period, and the government making a matching contribution equal to 18.5% (an increase of 4.5 percentage points compared to NPS) of salary for each month the employee is working. In contrast, the additional 8.77% number discussed in the previous paragraph is a one-time cost to the government. The proposed pension guarantee scheme discussed in the paper is also better for the retirees in the sense that the scheme guarantees at least 50% of inflation adjusted salary, and also lets the retirees keep any pension amount received in excess of the (inflation adjusted) 50% last drawn salary, unlike the UPS. One can also use the framework discussed in the paper to come up with numbers for the aggregate cost of implementing the pension guaranteed scheme. Such a calculation would require information on number of employees, along with their salary and years of service left. To see why information on years of service left is important, we note that the 8.77% number calculated in previous paragraph is based on 30 years of service period. Keeping other parameters same, this number becomes 0.58% if the employee has a service period of 35 years, and becomes 43.27% if the employee has a service period of 25 years instead. Increase in number of years of service leads to a reduction in the amount required to provide pension guarantee mainly due to the higher amounts of contribution to the corpus as well as higher time period for the corpus to compound. In future research, we plan to use detailed information about public sector employees' service periods and pay scales to calculate aggregate expenditure required to implement such a scheme and compare the cost with similar government schemes.

**Conclusion**

In this paper, we have developed a framework to analyse the risks associated with market returns and annuity rates at the time of retirement in a defined contribution system like the New Pension System (NPS). By using historical data on the performance of the benchmark index for central government employees, we have quantified the impact of these risks on the pension corpus. Our results show how fluctuations in market returns and annuity rates can leave individuals financially vulnerable if the corpus is insufficient to sustain them during retirement.

We have also calculated the potential cost to the government of hedging these risks by providing guarantees to individuals in the NPS. This analysis highlights the financial burden that the government may need to bear to ensure the stability and sustainability of such pension systems.

Moreover, the framework we have proposed is not limited to analysing market risks alone. It can be extended to study other significant risks faced by pension systems, such as inflation risk and longevity risk. These risks can further impact the adequacy of the pension corpus and the associated costs to the government. While this paper focuses on market risks, addressing these additional risks is crucial for developing a comprehensive understanding of the challenges in pension systems.

Future research can build upon this framework to include these broader risks, enabling policymakers to make informed decisions for creating a more secure and sustainable pension system. Our work provides a strong foundation for such efforts and underscores the importance of proactive risk management in pension design to ensure financial security for retirees and mitigate the fiscal burden on the government.